\documentclass[aps,prl,twocolumn,,superscriptaddress,floatfix]{revtex4-1}
\usepackage[svgnames]{xcolor}
\usepackage[colorlinks=true,urlcolor = purple,linkcolor=teal,citecolor=magenta,bookmarks=false]{hyperref}
\usepackage{amsfonts,amsmath,amssymb}
\usepackage{graphicx}
\usepackage{dsfont}

\usepackage{lipsum}
\usepackage{graphicx}
\usepackage{amsmath,amssymb}
\usepackage{braket}
\usepackage{mathtools}
\usepackage{hyperref}
\usepackage{multibib}

\usepackage{booktabs,amsmath}

\usepackage{color}
\usepackage[normalem]{ulem}
\usepackage{amsfonts}

\usepackage{pdfpages} 
\makeatletter
\AtBeginDocument{\let\LS@rot\@undefined}
\makeatother

\usepackage{amssymb}
\usepackage{pifont}
\newcommand{\cmark}{\ding{51}}%
\newcommand{\xmark}{\ding{55}}%

\definecolor{mypink1}{rgb}{0.858, 0.188, 0.478}
\definecolor{dartmouthgreen}{rgb}{0.05, 0.5, 0.06}
\hypersetup{
    colorlinks=true,
    linkcolor=blue,
    filecolor=blue,      
    urlcolor=mypink1,
    citecolor=dartmouthgreen,
}

\usepackage{xcolor,colortbl}

\definecolor{DGreen}{rgb}{0, 0.5, 0.0}

\usepackage{pdfpages} 
\makeatletter
\AtBeginDocument{\let\LS@rot\@undefined}
\makeatother

\begin{document}

\title{Stability of superdiffusion in nearly integrable spin chains}

\author{Jacopo De Nardis}
\affiliation{Laboratoire de Physique Théorique et Modélisation, CNRS UMR 8089,CY Cergy Paris Université, 95302 Cergy-Pontoise Cedex, France}
\author{Sarang Gopalakrishnan}
\affiliation{Department of Physics, The Pennsylvania State University, University Park, PA 16820, USA}
\author{Romain Vasseur}
\affiliation{Department of Physics, University of Massachusetts, Amherst, Massachusetts 01003, USA }
\author{Brayden Ware}
\affiliation{Department of Physics, University of Massachusetts, Amherst, Massachusetts 01003, USA }

\begin{abstract}

Superdiffusive finite-temperature transport has been recently observed in a variety of integrable 
systems with nonabelian global symmetries.
Superdiffusion is caused by giant Goldstone-like quasiparticles stabilized by integrability. 
Here, we argue that these giant quasiparticles remain long-lived, and give divergent contributions to the low-frequency conductivity $\sigma(\omega)$, even in systems that are not perfectly integrable. 
We find, perturbatively, that $ \sigma(\omega)  \sim \omega^{-1/3}$ for translation-invariant static perturbations that conserve energy, and $\sigma(\omega) \sim | \log \omega |$ for noisy perturbations. 
The (presumable) crossover to regular diffusion appears to lie beyond low-order perturbation theory.
%
By contrast, integrability-breaking perturbations that break the nonabelian symmetry yield conventional diffusion. 
Numerical evidence supports the distinction between these two classes of perturbations.

\end{abstract}

\maketitle
\textbf{\textit{Introduction}}. Many paradigmatic quantum models in one dimension are integrable, including the Heisenberg and Hubbard models; thus, \emph{approximate} integrability is ubiquitous in one-dimensional quantum materials, and strongly influences their response properties~\cite{bertini2020finite}. 
Although integrable models are in some sense ``exactly solvable,'' their finite-temperature and nonequilibrium dynamical properties have proved difficult to compute exactly. 
However, recent years have seen enormous progress in our understanding of the dynamics of integrable systems~\cite{PhysRevLett.98.050405,Rigol:2008kq}, due to the development of improved numerical techniques~\cite{PhysRevLett.106.220601,karraschdrude,2017arXiv170208894L, PhysRevB.97.035127, sanchez2018anomalous, PhysRevB.89.075139,1367-2630-15-8-083031,lzp, PhysRevLett.122.210602, dupont_moore,2020arXiv200405177R,2020arXiv201207849L}, as well as recent experiments~\cite{kinoshita,Langen207,tang2018,PhysRevLett.122.090601,Jepsen:2020aa,2020arXiv200608577M, 2020arXiv200906651M,2020arXiv200913535S}, and new theoretical methods such as generalized hydrodynamics (GHD)~\cite{Doyon, Fagotti, SciPostPhys.2.2.014, PhysRevLett.119.020602,  BBH0, BBH,PhysRevLett.119.020602, GHDII, doyon2017dynamics, solitongases,PhysRevLett.119.195301,2016arXiv160408434Z, PhysRevB.96.081118,PhysRevB.97.081111, PhysRevLett.120.164101, dbd1, ghkv, dbd2, gv_superdiffusion,Balasz, horvath2019euler, PhysRevB.100.035108,2019arXiv190601654B,10.21468/SciPostPhys.8.3.041,Biella:2019aa, ruggiero2019quantum,PhysRevLett.125.070602,PhysRevLett.125.240604}. 
Perhaps the most surprising outcome of these developments has been the discovery that superdiffusive spin transport~\cite{PhysRevLett.106.220601,lzp}, with the dynamical scaling exponent $z = \frac{3}{2}$ and scaling functions~\cite{PhysRevLett.122.210602} that appear to lie in the Kardar-Parisi-Zhang (KPZ) universality class~\cite{kpz}, is generic in integrable spin chains that are invariant under a continuous nonabelian symmetry~\cite{dupont_moore,PhysRevB.102.115121,2020arXiv200908425I}. Superdiffusive spin transport was also observed in recent inelastic neutron scattering experiments~\cite{2020arXiv200913535S}.
Although superdiffusion is not yet fully understood, it has become clear (from multiple lines of argument) that the degrees of freedom responsible for it are ``giant quasiparticles'' of the integrable model~\cite{idmp,gv_superdiffusion,PhysRevLett.123.186601,gvw,vir2019,dmki}, corresponding to large solitonic wavepackets made up of Goldstone modes~\cite{PhysRevLett.125.070601,2020arXiv200908425I}. These giant quasiparticles consist in local rotations of the nonabelian vacuum, and therefore cost little energy. Because these systems are integrable, giant quasiparticles are stable even at finite temperature, and therefore non-trivially influence spin and charge transport. 

The fate of superdiffusion when integrability is weakly broken is a question of great experimental relevance~\cite{2020arXiv200913535S}. Away from integrability, neither exact solutions nor infinitely stable quasiparticles exist; the theoretical tools we have to deal with this regime are still primitive~\cite{sirker:2010,PhysRevB.83.035115,PhysRevLett.115.180601,huangkarrasch,PhysRevX.9.021027,bulch2019superdiffusive,dmki,2020arXiv200713753G}. A natural framework to address weakly decaying quasiparticles is the Boltzmann equation~\cite{friedman2019diffusive,2020arXiv200411030D}, which relies on the matrix elements of generic local operators between eigenstates of the integrable system. These matrix elements are in general unknown, despite some recent progress~\cite{panfil2018,dbd2}; their asymptotic form is known in a few special limits, such as slowly fluctuating noise~\cite{bastianello2020generalised} or atom losses~\cite{10.21468/SciPostPhys.9.4.044,2020arXiv201215640H}. 
\begin{table}
    \begin{tabular}{| c | c | c |}
    \hline
 Symmetry & SU$(2)$ \textcolor{DGreen}{\cmark}  & SU$(2)$ \textcolor{red}{\xmark}  \\ \hline
  Energy \textcolor{DGreen}{\cmark}   & $\sigma(\omega) \sim \omega^{-1/3}$ &$ \sigma(\omega) \sim \sigma^{\rm d.c.}$ \\   
  Crystal momentum \textcolor{DGreen}{\cmark}  & & \\ \hline
  Energy \textcolor{red}{\xmark}  & $\sigma(\omega) \sim | \log \omega|$ &$ \sigma(\omega) \sim \sigma^{\rm d.c.}$ \\ 
  Crystal momentum \textcolor{DGreen}{\cmark} & & \\ \hline
  Crystal momentum \textcolor{red}{\xmark}  & $ \sigma(\omega) \sim \sigma^{\rm d.c.}$ & $ \sigma(\omega) \sim \sigma^{\rm d.c.}$ \\
  (disorder) &  & \\

  \hline
    
    \end{tabular} \label{Tab1}
    \caption{{\bf Perturbative predictions} for the a.c. spin conductivity of perturbed Heisenberg spin chains at intermediate frequencies, depending on whether the SU$(2)$ symmetry is preserved (\textcolor{DGreen}{\cmark}) or broken (\textcolor{red}{\xmark}).}
 \end{table}
In the present work we show that symmetry considerations constrain the effects of integrability breaking, and determine the fate of superdiffusion in the limit of weak integrability breaking. In particular, local perturbations that are invariant under the global nonabelian symmetry \emph{cannot} scatter giant quasiparticles, because these quasiparticles look locally like the vacuum. For a giant quasiparticle of size $\ell$, the fastest possible decay allowed by symmetry is $\ell^{-2}$; phase space constraints can suppress this further. As a consequence, \emph{within perturbation theory}, it appears that some form of anomalous diffusion survives: for symmetry-preserving noise, one has a logarithmically divergent diffusion constant, whereas for symmetry-preserving Hamiltonian perturbations, the KPZ scaling is unaffected by the perturbation to leading order. By contrast, perturbations that break the nonabelian symmetry can immediately dismember giant quasiparticles and restore diffusion. The distinction between these two types of behavior is clearly visible in our numerical studies of spin transport.

\textbf{\textit{Heisenberg spin chain}}. Although our analysis will extend to other isotropic integrable models, we will focus for concreteness on the perturbed Heisenberg spin chain  
\begin{equation} \label{eqHeisenberg}
H = H_0 + g V, \quad H_0 = J \sum_n \vec{S}_n \cdot \vec{S}_{n+1}.
\end{equation}
Here, $ \vec{S}_j$ are spin-$\frac{1}{2}$ operators on site $j$, $V$ is a possibly time-dependent integrability breaking perturbation with $g$ small, and we will set $J=1$ in the following.  
Spin transport in the integrable limit $g=0$ is known to be {\em superdiffusive}, while energy transport is purely ballistic. For simplicity, we will focus on the high-temperature regime, although our conclusions carry over to arbitrary finite temperatures. 
We will characterize spin transport by the a.c. spin conductivity, which at high temperature is given by the Kubo formula 
\begin{equation}\label{eq:kubo}
\sigma(\omega) = \beta \int_0^\infty dt \  {\rm e}^{i \omega t}  \langle J(t) j_0(0) \rangle_\beta 
\end{equation}
with the spin current $j_n = -i (S^+_n S^-_{n+1} - {\rm h.c.})$, $J = \sum_n j_n$, and with $ \langle \dots \rangle_\beta$ referring to equilibrium ensemble average at temperature $\beta^{-1}$. 

We first summarize the key ingredients responsible for spin superdiffusion in the model~\eqref{eqHeisenberg} in the integrable $g = 0$ limit. Here, the a.c. conductivity scales as $\sigma(\omega) \sim \omega^{-1/3}$, corresponding to a dynamical exponent $z=3/2$. This exponent can be understood by analyzing the quasiparticles of the Hamiltonian $H_0$. Those excitations are built starting from a reference ferromagnetic state (``vacuum'') and adding (non-trivially dressed) magnons and bound states thereof called ``strings.'' Strings are labelled by a quantum number $s = 1, 2, \dots$ (with $s=1$ corresponding to single magnons), and a continuous momentum $k_s$ that will play little role in our analysis. 
The thermodynamic properties of these quasiparticles are known exactly from the thermodynamic Bethe ansatz (TBA) solution of $H_0$: for example, their density in an equilibrium finite-temperature Gibbs state scales as $\rho_s \sim 1/s^3$.
At large $s$, they have been identified with macroscopically large solitons made out of interacting Goldstone modes (slow modulations of the vacuum orientation) above the ferromagnetic vacuum \cite{PhysRevLett.125.070601,vir2019}. 
While it might seem counter-intuitive that long-wavelength modes matter to high-temperature physics, their stability is protected by integrability. Their properties follow from ferromagnetic Goldstone mode physics: a string $s$ has width (size) $s$ and is made out of Goldstone modes with momentum $ \sim s^{-1}$, and thus have a small energy density $ \sim s^{-2}$ using the dynamical exponent $z=2$ of ferromagnetic Goldstone modes. The corresponding velocity is $v_s \sim 1/s$, so large-$s$ strings move slowly. The energy carried by each string is suppressed as $\varepsilon_s  
\sim s^{-1}$ (consistent with the intuition that they are ``soft'' excitations), so we immediately see that energy transport in this model is ballistic (since the quasiparticles have a finite velocity), and is dominated by small strings. 

\textbf{\textit{Superdiffusion in the integrable limit}}. We now turn to spin transport, which is dominated by large strings instead. We give an argument for spin superdiffusion that is reformulated from Refs.~\cite{gv_superdiffusion,gvw} in a way that is physically more transparent.
When propagating in vacuum, $s$-strings carry a spin $m_s=s$ (their number of magnons). In thermal states, strings are ``screened'' via non-perturbative dressing due to interaction effects, and effectively become neutral~\cite{PhysRevLett.78.943,PhysRevB.57.8307}; their dressed, or effective, magnetization vanishes $m^{\rm dr}_s = 0$ in any thermal state with no net magnetization. Specifically, an $s$-string is screened when it encounters an $s' > s$-string (just as a magnon is screened when it passes through a domain wall~\cite{gv_superdiffusion}); such collisions set the ``lifetime'' for an $s$-string to transport magnetization. The density of strings bigger than $s$ is $\rho_{s'>s}=\sum_{s' >s} \rho_s \sim 1/s^2$: thus, an $s$-string moving at speed $v_s \sim 1/s$ first encounters a larger string on a timescale $\tau^0_s \sim (1/\rho_{s'>s})/v_s \sim s^3$. This characteristic timescale was identified using different approaches in previous works~\cite{gv_superdiffusion,gvw,2019arXiv190905263A}, and underlies the physics of superdiffusion. At a given time $t$, such that $t \gg \tau^0_s$, small strings are screened, namely $m^{\rm dr}_s =0$, and do not contribute to transport. As a result, as time increases, spin transport is dominated by strings with larger and larger $s$, namely such that $ \tau^0_s \sim s^3 \sim t$.

We now apply this logic to the Kubo formula \eqref{eq:kubo}. At a time $t$ inside the integral in \eqref{eq:kubo}, strings with $s \leq t^{1/3}$ have been completely screened and carry no net current, whereas those with $s \geq t^{1/3}$ still carry their original current $j_s \sim  v_s m_s ={\cal O}(1)$ where we used $v_s  m_s \sim s^{-1} \times s$. As the screening gives rise to exponential decay of their contributions to spin current~\footnote{The probability of avoiding collisions with larger strings decreases exponentially with time.},  one can write $\langle J(t) j_0(0) \rangle \sim \sum_s \rho_s (m_s v_s)^2 e^{-t/\tau_s^0}$. Plugging into the Kubo formula, 
\begin{equation} \label{eqKuboIntegrable}
\sigma(\omega) \sim \int_0^\infty dt \: {\rm e}^{i \omega t} \sum_{s \geq 1} s^{-3}  {\rm e}^{-t/s^3} \sim \omega^{-1/3}. 
\end{equation}
As expected, the conductivity diverges at low frequency with an exponent corresponding to $z=3/2$. The precise form of the exponential cutoff~\eqref{eqKuboIntegrable} is not important, and can be replaced with any integrable function of $t/s^3$. However, eq.~\eqref{eqKuboIntegrable} has an appealing physical interpretation where each string gives a Lorentzian (diffusive) contribution to the conductivity of width $1/\tau^0_s \sim s^3$. 

\begin{figure*}[!t]
	\includegraphics[width=.95\textwidth]{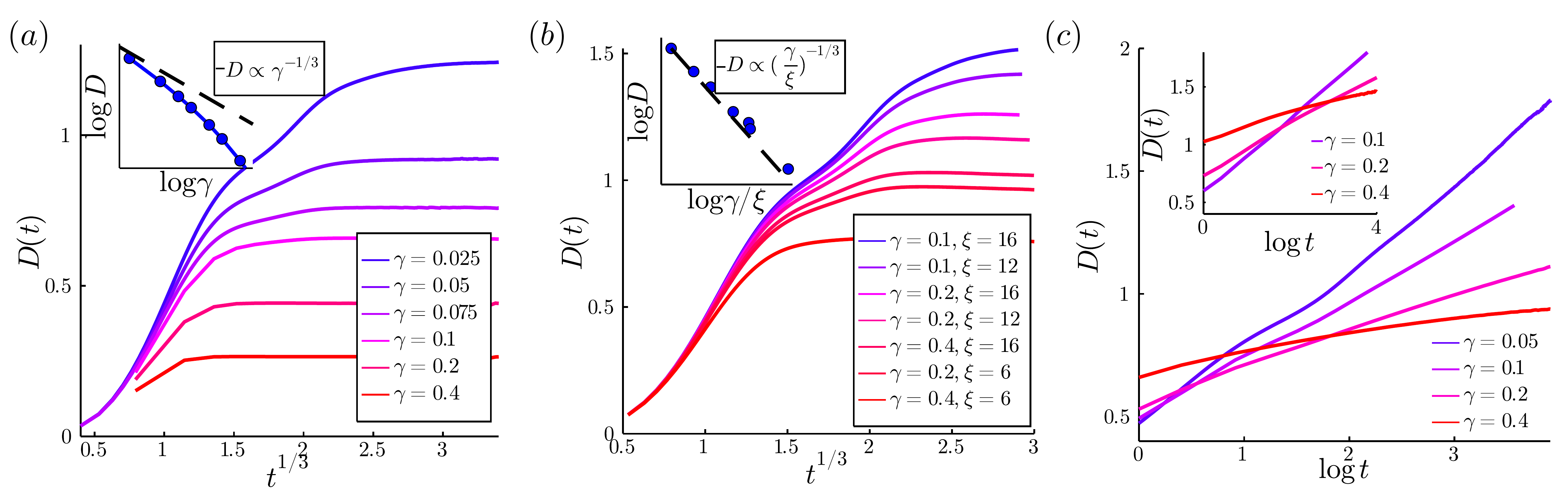}
	\caption{{\bf Noisy perturbations}. (a)~Apparent diffusion constant $D(t)$ vs. time $t$, for noise coupling to $S^z_i$ with spatially uncorrelated fluctuations. The diffusion constant rapidly saturates to a value plotted in the inset vs. noise strength $\gamma \sim g^2$. (b)~Same as (a), but for noise with a spatial correlation length $\xi$. Inset: Scaling of the diffusion constant compared to the prediction $D \sim (\gamma/\xi)^{-1/3}$. (c)~Apparent logarithmic divergence of $D(t)$ vs. $t$ for noise that preserves SU$(2)$ symmetry: both for noise coupling to the energy density (main panel) and for noise coupling to the energy current density (inset).}
	\label{fig1}
\end{figure*}

\textbf{\textit{Quasiparticle lifetimes}}. We now consider the effects of (small) integrability-breaking perturbations $g \neq 0$, which preserve the total $z$-component of the magnetization $S^z = \sum_n S^z_n$ and may preserve the energy, but break all other conservation laws.
We incorporate integrability-breaking terms by writing a Boltzmann equation for the quasiparticles, with a collision integral---incorporating quasiparticle scattering and decay---computed perturbatively using Fermi's Golden Rule~\cite{friedman2019diffusive,2020arXiv200411030D}. This approach is valid in the ``Boltzmann regime'' at long times and small $g$, with $t g^2$ fixed; we will discuss the roles of higher-order terms in perturbation theory below. At the level of our analysis, the main effect of the perturbation $V$ is to give a new, finite lifetime $\tau_s$, with decay rates $\Gamma_s = \tau_s^{-1}$ of order ${\cal O}(g^2)$. Starting from a representative thermal state $\left| \rho \rangle \right.$, containing all strings with thermal densities $\rho_s$,
those decay rates are obtained by summing over all possible accessible states $\left| n \rangle \right.$ compatible with the residual conservation laws (energy and/or quasi-momentum conservation), with a rate given by the square of the matrix element $| \langle n |V | \rho \rangle  |^2$ with the appropriate density of state factors. The a.c. spin conductivity is then given by:
\begin{equation} \label{eqKuboNonIntegrable}
\sigma(\omega) \sim \int_0^\infty dt \: {\rm e}^{i \omega t} \sum_{s \geq 1} s^{-3} {\rm e}^{-t/\tau_s^0} \:  {\rm e}^{-  t \Gamma_s}. 
\end{equation}
%
As we will see below, different classes of perturbations lead to very different scalings of the rates $\Gamma_s$ with $s$, with some perturbations allowing for long-lived giant strings. 

\textbf{\textit{Symmetry-breaking noisy perturbations}}. 
We first consider a generic perturbation that breaks the spin-rotation symmetry. In this case, we expect the matrix element of the perturbation to be either ${\cal O}(1)$, or possibly to increase with $s$ (since for example, the charge of an $s$-string is $m_s=s$). 
If we further assume that the integrability breaking is due to temporally fluctuating noise, then the density of available states is clearly non-vanishing, since an $s$-string can scatter by changing its momentum. (This might not be the dominant process, of course.)
In this case, $\Gamma_s \sim s^\alpha$ with $\alpha>0$ in eq.~\eqref{eqKuboNonIntegrable} leading to a finite d.c. value $\sigma^{\rm d.c.} = \lim_{\omega \to 0} \sigma(\omega)$. 
This is the expected behavior of integrability breaking, which generically should lead to diffusive transport for residual conserved charges. Note that eq.~\eqref{eqKuboNonIntegrable} predicts that the finite time diffusion constant $D(t) = \frac{\beta}{\chi} \int_0^t \sum_n \langle j_n(t) j_0(0) \rangle_\beta$, with $\chi$ the spin susceptibility,  approaches its asymptotic value very quickly, with exponential convergence in time. While anomalous transport is washed out by the integrability-breaking perturbation in this case, remnants of the anomalous exponent $z=3/2$ can be observed in the dependence of the diffusion constant on the integrability-breaking parameter $g$. If we convolve the anomalous integrable scaling $\omega^{-1/3}$ by a Lorentzian of width $\Gamma \sim g^2$, we immediately find 
\begin{equation} \label{eqScalingDiffusion}
\sigma^{\rm d.c.} = \chi D \sim g^{-2/3}.
\end{equation}
To leading-order, the small $g$-dependence of the susceptibility can be ignored, and we see that the diffusion constant scales in a non-analytic way with $g$. Similar non-analytic dependences of diffusion constants were reported in Refs.~\cite{friedman2019diffusive,PhysRevLett.125.180605}.

To check these predictions, we consider a noisy perturbation $V(t) = \sum_n \eta_n (t) S^z_n$, with $\eta_n $ some classical correlated noise $\langle \eta_n(t) \eta_{n^\prime}(t^\prime) \rangle = \gamma \delta(t-t^\prime) f(n-n')$, and $\gamma \sim g^2$ the strength of the noise. {Such correlated noise 
can be expected to model generic external perturbations at large scales.} We implemented time evolution using a Lindblad approach and matrix product operators (MPO)~\cite{suppmat}. We considered both uncorrelated ($f(n) = \delta_{n,0}$) and correlated ($f(n) \sim {\rm e}^{-|n| /\xi}$) noisy perturbations. Our results are consistent with diffusive transport and a diffusion constant scaling as $D \sim (\gamma/\xi)^{-1/3}$, as expected from~\eqref{eqScalingDiffusion} (Fig.~\ref{fig1}), and with a relaxation time $\sim 1/\gamma$~\cite{suppmat}.

\textbf{\textit{Symmetric noisy perturbations}}. We now turn to a much more interesting class of perturbations that preserve the spin-rotation SU$(2)$ symmetry of $H_0$. Intuitively, the action of such perturbations on large strings should be suppressed since those are smooth vacuum rotations. In fact, Goldstone physics implies that the matrix elements of any SU$(2)$-invariant must be suppressed with $s$ at least as energy is, as $\varepsilon_s \sim 1/s$. The matrix element for large strings to decay is thus suppressed, so we can expect large strings to be long-lived, leading potentially to anomalous transport. 

In order to compute the scaling of the decay rates $\Gamma_s$, we need to consider the processes involved and the accessible density of states more carefully. For concreteness, we first consider the case of noise coupling to energy:  $V(t) = \sum_n \eta_n (t) \vec{S}_n \cdot \vec{S}_{n+1} $ with $\eta_n$ some uncorrelated white noise as before. As this perturbation breaks both energy and momentum conservation, the leading processes will be single particle-hole excitations where a given string$s$ with momentum $k_s$ will scatter into another mode with new momentum $k_s'$~\cite{friedman2019diffusive,bastianello2020generalised} (causing rapidity or momentum diffusion~\cite{bastianello2020generalised}). Strings $s$ can also decay into multiple smaller strings while preserving $S^z$. Now the size of the Brillouin zone for an $s$-string goes as $\sim 2\pi/s$, as $k_s \sim s^{-1}$, but the noise perturbation allows for processes scattering across different Brillouin zones, therefore the accessible density of states for a single particle-hole process does not scale with $s$~\cite{suppmat}. As a result, the decay rates for an $s$-string is set solely by the matrix element of the perturbation, and scales as its square, namely $\Gamma_s \sim |\langle k_s | V | k'_s \rangle|^2 \sim  1/s^2$. The late scaling can be deduced by the known low-energy limit of the matrix element $\lim_{k_s' \to k_s}\langle k_s | V | k'_s \rangle = \varepsilon^{\rm dr}_s \sim s^{-1}$~\cite{dbd2,doyon2019lecture}, where $\varepsilon^{\rm dr}_s$ is the dressed energy of the string.  Giant strings are long lived, but decay before they get screened. As this order in perturbation theory, this leads to a log-divergence of $\sigma(\omega)$:
\begin{equation}
\sigma(\omega) \sim \sum_{s} \frac{1}{s^3} \frac{s^{-2}}{\omega^2 + s^{-4}} \sim| \log \omega|,
\end{equation}
{which can be seen by noticing that the sum is approximately given by $ \sum_{s \geq 1}^\Lambda \frac{1}{s} $ for $\Lambda \sim 1/\omega^{1/2} $.} 
Remarkably, giant strings are long-lived enough to lead to superdiffusive transport, albeit in the weaker form of logarithmic corrections to diffusion. Similar log-diffusion scalings have recently been observed  up to long times in various isotropic spin chains, and was interpreted in the framework of the KPZ equation in Ref.~\cite{dmki}, though the relation to integrability has remained controversial~\cite{2020arXiv200713753G}.  

 We have checked this prediction using MPO simulations by considering uncorrelated noise coupling to either energy density $\vec{S}_n . \vec{S}_{n+1}$, or energy current density  $\vec{S}_n . (\vec{S}_{n+1} \times \vec{S}_{n+2})$ (Fig.~\ref{fig1}). In both cases, our results are consistent with logarithmic diffusion $D(t) \sim \log t$ for weak enough perturbations. For stronger perturbations we do observe a trend towards saturation in $D(t)$, indicating that higher-order processes in perturbation theory or non-perturbative processes can eventually restore ordinary diffusion.
 
\begin{figure}[t!]
	\includegraphics[width=1.0\columnwidth]{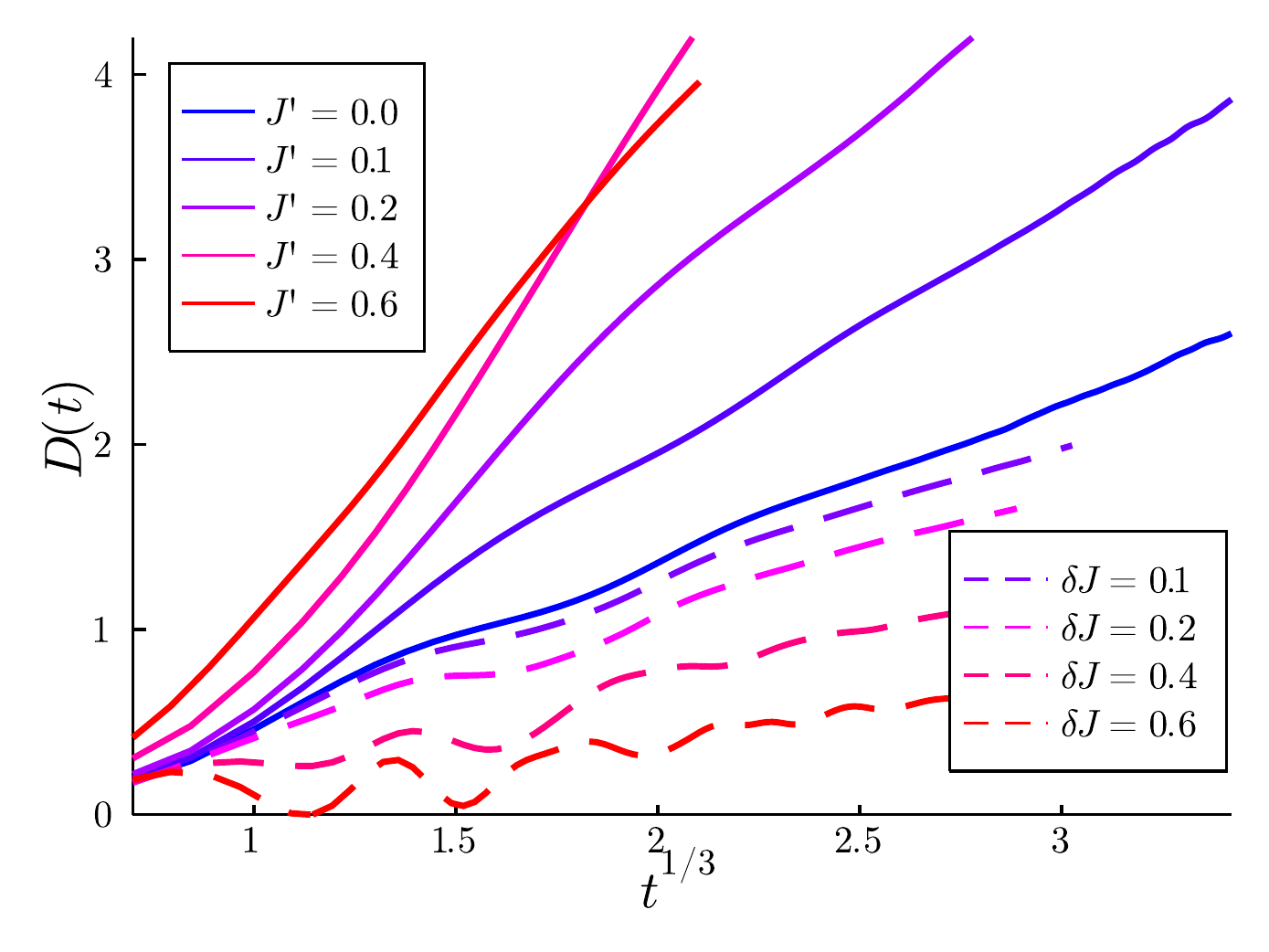}
	\caption{{\bf Static perturbations}. Time-dependent diffusion constant $D(t)$ vs. $t^{1/3}$ for next-nearest neighbor couplings ($g V =  J' \sum_n \vec{S}_{n} \cdot \vec{S}_{n+2}$, solid lines) and staggered fields ($g V = \delta J \sum_n (-1)^n \vec{S}_{n} \cdot \vec{S}_{n+1}$, dashed lines) of various strengths. The KPZ behavior $z=3/2$ seems to persist in all cases.}
	\label{fig2}
\end{figure}

\textbf{\textit{Static perturbations}}. Finally, we consider static, translation-invariant SU$(2)$ preserving perturbations. The argument about suppressed matrix elements for large strings still applies, but the main difference with the noisy perturbation considered above is that energy and (quasi)momentum conservation greatly constrains the allowed processes contributing to $\Gamma_s$. In particular, single particle-hole processes are not allowed. Because of the dispersion mismatch between different strings, two-particle umklapp scattering processes do not seem capable of relaxing momentum either. Thus, relaxation occurs by scattering processes with three or more of them. 
While we have little analytic control over such processes, we find numerically that the superdiffusive behavior with time-dependent diffusion constant $D(t) \sim t^{1/3}$ ($z=3/2$) persists up to all the time scales we access numerically, even for sizable perturbations (Fig.~\ref{fig2}).
This suggests that the decay rate falls off as $\Gamma_s \sim 1/s^3$ (or faster), so that decay is no longer parametrically faster than screening. The bound $\Gamma_s \leq 1/s^2$ from Goldstone physics continues to hold, but appears not to be saturated in this case.
%
%
As a result, we find that the anomalous scale $z=3/2$ is remarkably robust to isotropic integrability-breaking perturbations. We do expect that at long enough times, higher-order processes in perturbation theory will eventually take over and lead to regular diffusion, though we were not able to access this crossover regime numerically. We refer the reader to the supplemental material for additional numerical results for other perturbations breaking either translation invariance (disorder) or SU$(2)$~\cite{suppmat}. {Our final results are summarized in Table~I.}


\textbf{\textit{Discussion}}. 
We have discussed the fate of anomalous spin diffusion in spin chains with nonabelian symmetries in the presence of weak integrability-breaking perturbations. Since anomalous transport is due to thermally dressed ``Goldstone solitons,'' its fate depends on the symmetries of the integrability-breaking perturbation. When the perturbation breaks the nonabelian symmetry, the Goldstone solitons are immediately unstable and diffusion immediately sets in. When the perturbation preserves the symmetry, however, it cannot effectively scatter Goldstone solitons, so some form of anomalous diffusion persists. We have argued that this should give rise to a diffusion constant that diverges at least logarithmically at low frequencies and perhaps faster, depending on phase-space constraints. These symmetry considerations directly extend to other Lie-group symmetry. 
 While we expect---and our numerics suggests---that regular diffusion is eventually restored on some long timescale, the dependence of this timescale on the integrability-breaking parameter $g$ lies outside the scope of low-order perturbation theory. Identifying the relevant decay channels and understanding the crossover to regular diffusion is an interesting task for future work. Another interesting question is whether the mechanisms explored here can manifest themselves, e.g., as anomalously large long-time tails in classical hydrodynamics, as suggested in Ref.~\cite{2020arXiv200713753G}.

\begin{acknowledgments}

{\textit{\textbf{Acknowledgments}}}. 
We are grateful to L. Delacretaz, P. Glorioso, E. Ilievski, V. Khemani, T. Rakovszky, and C. Von Keyserlingk for helpful discussions. S.G. acknowledges support from NSF DMR-1653271. R.V. acknowledges support from the Air Force Office of Scientific Research under Grant No. FA9550-21-1-0123, and the Alfred P. Sloan Foundation through a Sloan Research Fellowship.

\end{acknowledgments}

\bibliography{refs}

\bigskip

\onecolumngrid
\newpage



\section*{Supplementary material}

# Supplemental Material: Stability of superdiffusion in nearly integrable spin chains

Jacopo De Nardis[1], Sarang Gopalakrishnan[2], Romain Vasseur[3], and Brayden Ware[3]

[1] *Laboratoire de Physique Théorique et Modélisation, CNRS UMR 8089,*
*CY Cergy Paris Université, 95302 Cergy-Pontoise Cedex, France*
[2] *Department of Physics, The Pennsylvania State University, University Park, PA 16820, USA*
[3] *Department of Physics, University of Massachusetts, Amherst, MA 01003, USA*
(Dated: June 21, 2021)

## I. DETAILS OF THE MATRIX PRODUCT OPERATOR (MPO) CALCULATIONS

We numerically simulate real time dynamics of quantum systems with Markovian noise by evolving operators with the Lindblad master equation

$$\dot{\rho} = \mathcal{L}[\rho] = -i[H,\rho] + \sum_\mu \left( L_\mu \rho L_\mu^\dagger - \frac{1}{2} L_\mu L_\mu^\dagger \rho - \frac{1}{2} \rho L_\mu L_\mu^\dagger \right).$$

We employ a matrix product representation of operators, which can be time evolved with straightforward generalizations of time-dependent DMRG employed for dynamics of unitary systems [S1, S2]. Rather than evolving the density matrix $\rho$ directly, we evolve time-dependent local operators $A(t)$ with the Heisenberg equations of motion $\dot{A} = \mathcal{L}^*[A]$ [S3]. We employ a Trotterization schemes for the superoperator $\mathcal{L}$ when it can be decomposed into sums of nearest neighbor terms or in terms acting on three consecutive sites [S4]. The evolutions are done by representing the operator $A(t)$ as an MPO and evolving with TDMRG. At each update of the MPO tensors, a truncated SVD is done, with the number of kept states being chosen to give a truncation error smaller than $\varepsilon = 10^{-12}$, up to a maximum of $\chi_{max} = 512$. When the number of states needed to limit the error to $\varepsilon$ rises above $\chi_{max}$, the evolution continues using only $\chi_{max}$ states. We compare the data to evolutions with smaller $\chi_{max}$ and larger $\varepsilon$ to ensure that it is converged.

### A. Unperturbed Heisenberg chain

First, we simulated real time dynamics of the isotropic Heisenberg chain

$$H_0 = \sum_j \vec{S}_j \cdot \vec{S}_{j+1}.$$

The correlation functions

$$C(j - j', t) = \langle \sigma_j^z(t) \sigma_{j'}^z(0) \rangle = \langle \sigma_j^z(t/2) \sigma_{j'}^z(-t/2) \rangle$$

are computed by MPO time evolution, taking advantage of time translation and time reversal symmetry. The evolution is done with a fourth order Trotterization with a time step $\delta t = 0.1$. We plot the time dependent diffusion constant, defined as

$$D(t) = \frac{1}{2} \frac{d}{dt} V(t), \text{ with } V(t) = \sum_j j^2 C(j,t) - \left( \sum_j j C(j,t) \right)^2.$$

The results, shown in Fig. S1, provide an match to the predicted asymptotic scaling from Ref. S5:

$$D(t) \sim \frac{2}{3} \left( \frac{10\pi}{27} \right)^{4/3} t^{1/3} \approx 0.816 t^{1/3}.$$

Linearly fitting the TDMRG computation of $D(t)$ versus $t^{1/3}$ matches the predicted slope to within one percent.

As a measure of the computational difficulty of the evolution, we use the operator entanglement entropy $S(t)$ [S6]. $S(t)$ is the entanglement of the MPO $\sigma^z(t)$ reinterpreted as a MPS for the bipartition between the left and right of the system. The growth of $S(t)$ necessitates a growing bond dimension to accurately capture the entire operator $\sigma^z(t)$,

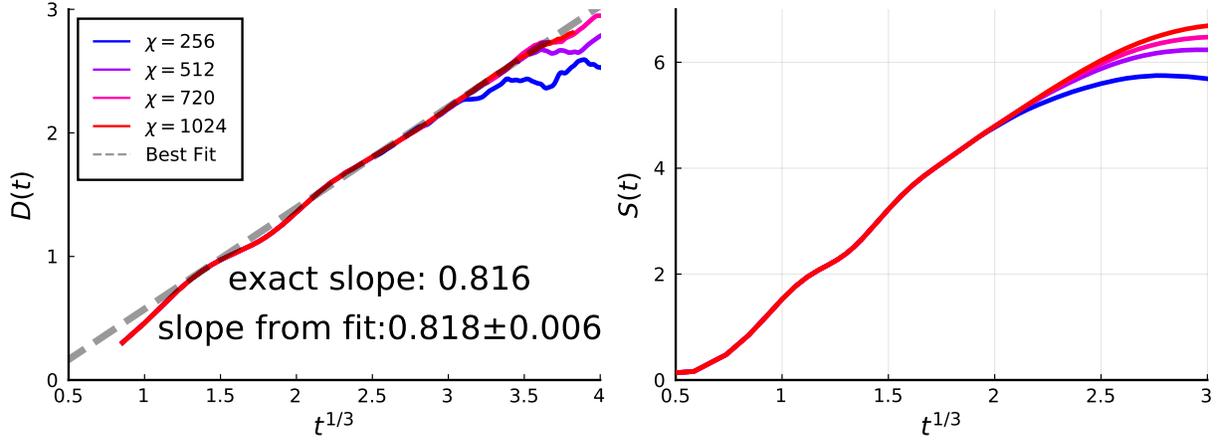

FIG. S1. **XXX Hamiltonian** Heisenberg chain dynamics simulated with a variety of bond dimensions $\chi_{\max}$. (left) Time-dependent diffusion constant scales asymptotically like $t^{1/3}$. (right) Operator entanglement entropy in the MPO evolution. It converges only for relatively short times with accessible bond dimensions.

as the operator entanglement entropy for an MPO of bond dimension $\chi$ satisfies the bound $S(t) \leq \log_2 \chi$. As we are evolving with fixed $\chi$, the convergence of $S(t)$ can only be achieved for relatively short times $t \lesssim 15$ with the bond dimensions of our computations. Our data is consistent with $S(t) \sim t^{1/3}$, as shown in Fig. S1, but due to the short time scale of the converged data we cannot rule out other forms for the asymptotic scaling of $S(t)$. Particularly, the scaling $S(t) \sim \log t$ was established for operator evolutions in free fermions and certain interacting integrable chains and has been conjectured to be a general result in integrable systems; our data is consistent with this conjecture as well [S7–S9].

### B. Correlated noise evolutions

We consider the isotropic Heisenberg spin chain $H_0$ with the addition of Markovian (uncorrelated in time) noise. The system is described by a stochastic Schrodinger equation $d\,|\psi\rangle\,/dt = -iH_\eta\,|\psi\rangle$, with

$$H_\eta = H_0 + \sum_j \eta_j(t)O_j.$$

The noise is generated by the correlated stochastic processes $\eta_j$ coupled to Hermitian operators $O_j = \sigma_j^z$. The noise is described by its correlations

$$\langle \eta_j(t)\eta_{j'}(t')\rangle = \gamma f(j - j')\delta(t - t'),$$

where $\gamma$ parameterizes the strength of the noise and $f(j)$ parameterizes the spatial correlations. This setup leads to the evolution equation

$$\frac{d\rho}{dt} = \mathcal{L}(\rho) = -i[H_0, \rho] - \frac{\gamma}{2}\sum_{jj'} f(j - j')(O_jO_{j'}\rho - O_j\rho O_{j'} - O_{j'}\rho O_j + \rho O_{j'}O_j),$$

for the noise averaged density matrix [S10]. As the $O_j$ commute for the $\sigma^z$ noise considered in this paper, the Liouvillian $\mathcal{L}$ can be written as

$$\mathcal{L} = -i\overrightarrow{H_0} + i\overleftarrow{H_0} - \frac{\gamma}{2}\sum_{jj'} f(j - j')(\overrightarrow{O_j} - \overleftarrow{O_j})(\overrightarrow{O_{j'}} - \overleftarrow{O_{j'}}),$$

where $\overrightarrow{O} = O \otimes \mathbf{1}$ and $\overleftarrow{O} = \mathbf{1} \otimes O$ are the superoperators corresponding to $O$ acting from the left or right, respectively. We work in the Heisenberg picture where the density matrix is static and operators evolve as [S3]

$$\frac{dA}{dt} = \mathcal{L}^*(A), \quad \mathcal{L}^* = i\overrightarrow{H_0} - i\overleftarrow{H_0} - \frac{\gamma}{2}\sum_{jj'} f(j - j')(\overrightarrow{O_j} - \overleftarrow{O_j})(\overrightarrow{O_{j'}} - \overleftarrow{O_{j'}}). \tag{S1}$$

We Trotter decompose the operator evolution

$$A(t + \delta t) = e^{\mathcal{L}^* \delta t} A(t) \approx e^{\mathcal{L}_1^* \delta t} e^{\mathcal{L}_0^* \delta t} A(t), \tag{S2}$$

where $\mathcal{L}_0$ and $\mathcal{L}_1$ correspond to the evolutions by $H_0$ and the noise respectively. The evolution by $H_0$ is further decomposed using a standard fourth order Trotter scheme. To evolve by $e^{\mathcal{L}_1^* \delta t}$, we first write $\mathcal{L}_1$ as an matrix product operator (MPO). Such a representation can be generated exactly for functions $f(j)$ that can be written as sums of decaying exponentials, or approximately for other functions $f(j)$ [S11]. We use a single decaying exponential $f(j) = e^{-|j|/\xi}$, where the corresponding MPO takes the bond dimension 3 form

$$
\begin{aligned}
M_{[1]} &= \begin{pmatrix} \mathbb{1} & -\gamma e^{-1/\xi}(\overleftarrow{O_1} - \overrightarrow{O_1}) & -\gamma/2(\overleftarrow{O_1} - \overrightarrow{O_1})^2 \end{pmatrix}, \\
M_{[j]} &= \begin{pmatrix} \mathbb{1} & -\gamma e^{-1/\xi}(\overleftarrow{O_j} - \overrightarrow{O_j}) & -\gamma/2(\overleftarrow{O_j} - \overrightarrow{O_j})^2 \\ 0 & e^{-1/\xi}\mathbb{1} & \overleftarrow{O_j} - \overrightarrow{O_j} \\ 0 & 0 & \mathbb{1} \end{pmatrix}, \quad 1 < j < L, \\
M_{[L]} &= \begin{pmatrix} -\gamma/2(\overleftarrow{O_L} - \overrightarrow{O_L})^2 \\ \overleftarrow{O_L} - \overrightarrow{O_L} \\ \mathbb{1} \end{pmatrix}.
\end{aligned}
\tag{S3}
$$

Given this MPO representation for $\mathcal{L}_1$, an approximation $W^I(\delta t)$ to the evolution operator $e^{\mathcal{L}_1^* \delta t}$ is generated using the $W^I$ scheme [S12]. This results in a MPO of bond dimension 2 consisting of the following tensors:

$$
\begin{aligned}
W^I_{[1]}(\delta t) &= \begin{pmatrix} \mathbb{1} - \gamma \delta t/2(\overleftarrow{O_1} - \overrightarrow{O_1})^2 & -\gamma \delta t e^{-1/\xi}(\overleftarrow{O_1} - \overrightarrow{O_1}) \end{pmatrix}, \\
W^I_{[j]}(\delta t) &= \begin{pmatrix} \mathbb{1} - \gamma \delta t/2(\overleftarrow{O_j} - \overrightarrow{O_j})^2 & -\gamma \delta t e^{-1/\xi}(\overleftarrow{O_j} - \overrightarrow{O_j}) \\ \overleftarrow{O_j} - \overrightarrow{O_j} & e^{-1/\xi}\mathbb{1} \end{pmatrix}, \quad 1 < j < L, \\
W^I_{[L]}(\delta t) &= \begin{pmatrix} \mathbb{1} - \gamma \delta t/2(\overleftarrow{O_L} - \overrightarrow{O_L})^2 \\ \overleftarrow{O_L} - \overrightarrow{O_L} \end{pmatrix}.
\end{aligned}
\tag{S4}
$$

The evolution by $W^I$ is a $1^{st}$ order approximation. To reduce the Trotter error, we construct the MPO $W(\delta t) = W^I(\frac{1+i}{2}\delta t)W^I(\frac{1-i}{2}\delta t)$, which instead is a $2^{nd}$ order approximation. To further reduce the Trotter error for the noisy part of the evolution, we additionally use a smaller time step repeated several times for each fourth order time step of the Hamiltonian evolution: $e^{\mathcal{L}_1^* \delta t} = \prod_{i=1}^n e^{\mathcal{L}_1^* \delta t/n} \approx \prod_{i=1}^n W(\delta t/n)$. For the numerics in this paper, $\delta t = 0.1$ and $n = 10$. By comparison with simulations for larger time steps, the results are converged in $\delta t$ and $n$.

## C. Uncorrelated noise evolutions

Uncorrelated noise can be reproduced using the evolution as in Eqs. (S1)-(S2) by taking $f(j - j') = \delta_{j,j'}$. However, we use a different parameterization of the noisy evolution which is more convenient for noise coupled to the multi-site operators $O_j = S_j \cdot S_{j+1}$ and $O_j = S_j \cdot (S_{j+1} \times S_{j+2})$. The noisy evolution is instead represented as a quantum channel with Kraus operators $M_a$, where the evolution of an operator is

$$A(t + \delta t) = \sum_a M_a^\dagger(\delta t) A(t) M_a(\delta t), \quad \sum_a M_a M_a^\dagger = \sum_a M_a^\dagger M_a = \mathbb{1}.$$

For $\sigma^z$ noise, the operators are taken to be

$$
\begin{aligned}
M_{1,j} &= \sqrt{1 - \gamma \delta t} \cdot \mathbb{I}, \\
M_{2,j} &= \sqrt{\gamma \delta t} \cdot \sigma_j^z.
\end{aligned}
$$

The evolution can then be represented as a product superoperator:

$$e^{\mathcal{L}_1 \delta t} = \prod_j \left( \sum_a \overrightarrow{M_{a,j}} \overleftarrow{M_{a,j}^\dagger} \right).$$

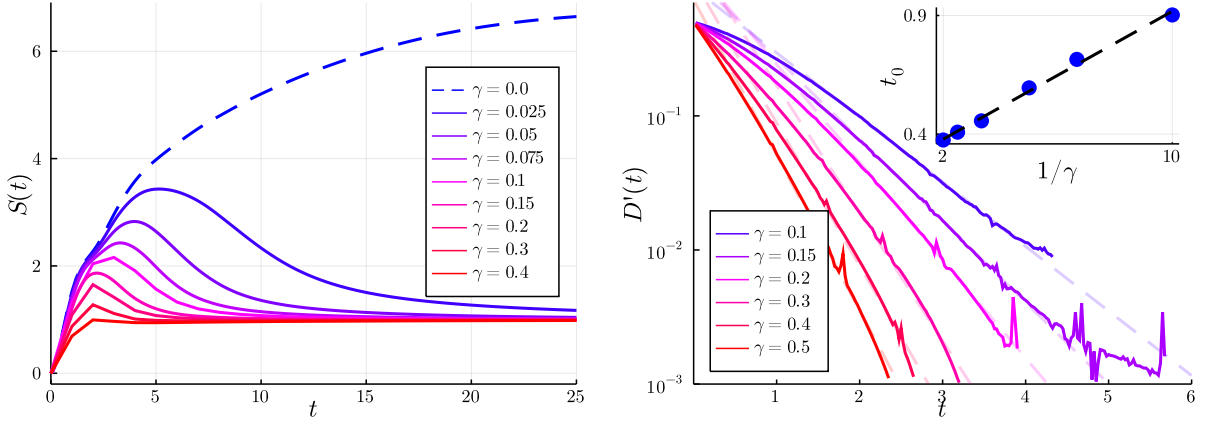

FIG. S2. **XXX Hamiltonian with uncorrelated $\sigma^z$ noise** (left) With non-zero $\sigma^z$ noise, the complexity of the simulation reaches a maximum in finite time and then drops — as illustrated here by the operator entanglement for the central cut. (right) $D(t)$ approaches the infinite time diffusion constant exponentially. This occurs on a time scale $t_0 \propto 1/\gamma$ (right inset).

For $S \cdot S$ noise, the operators are taken to be

$$M_{1,j} = \sqrt{1 - \gamma \delta t} \cdot \mathbb{I},$$

$$M_{2,j} = \sqrt{\gamma \delta t} \cdot (2 S_j \cdot S_{j+1} + \frac{1}{2}\mathbb{I}).$$

The evolution can then be represented as a superoperator circuit with two layers per time step:

$$e^{\mathcal{L}_1 \delta t} = \prod_{j \text{ even}} \left( \sum_a \overrightarrow{M_{a,j}} \overleftarrow{M_{a,j}^\dagger} \right) \prod_{j \text{ odd}} \left( \sum_a \overrightarrow{M_{a,j}} \overleftarrow{M_{a,j}^\dagger} \right).$$

For $S \cdot (S \times S)$ noise, the operators are taken to be

$$M_{1,j} = \sqrt{1 - \gamma \delta t} \cdot P_{\frac{1}{2}} + (\mathbb{1} - P_{\frac{1}{2}}),$$

$$M_{2,j} = \sqrt{\gamma \delta t} \left( \frac{4i}{\sqrt{3}} S_j \cdot (S_{j+1} \times S_{j+2}) \right)$$

where $P_{\frac{1}{2}}$ is the projector onto the spin-$\frac{1}{2}$ subspaces of the three consecutive spin-$\frac{1}{2}$s and satisfies

$$P_{\frac{1}{2}} = \left( \frac{4i}{\sqrt{3}} S_j \cdot (S_{j+1} \times S_{j+2}) \right)^2.$$

The evolution can then be represented as a superoperator circuit with three layers per time step:

$$e^{\mathcal{L}_1 \delta t} = \prod_{j=3,6,\ldots} \left( \sum_a \overrightarrow{M_{a,j}} \overleftarrow{M_{a,j}^\dagger} \right) \prod_{j=2,5,\ldots} \left( \sum_a \overrightarrow{M_{a,j}} \overleftarrow{M_{a,j}^\dagger} \right) \prod_{j=1,4,\ldots} \left( \sum_a \overrightarrow{M_{a,j}} \overleftarrow{M_{a,j}^\dagger} \right).$$

As in the previous section, the evolution is done by representing the operator $A(t)$ as an MPO and evolving with TDMRG. Each time step consists of a time step of the Hamiltonian evolution followed by a time step of the noisy evolution. The evolutions shown use a time step $\delta t = 0.1$, a truncation error of $\varepsilon = 10^{-12}$, and a maximum bond dimension of $\chi = 512$, which was sufficient to reach convergence for the shown time scales. Fig. 1 (a) (main text) and Fig. S2 show the results of the operator evolution of $\sigma^z(t)$ for a variety of noise levels $\gamma$. The symmetry breaking perturbation leads to diffusive transport, with the time dependent diffusion constant $D(t)$ quickly approaching its asymptotic value in time $1/\gamma$. The computational complexity of these evolutions is much smaller than those of the unperturbed integrable system, as the noise supresses the growth of the operator entanglement entropy $S(t)$ at long times.

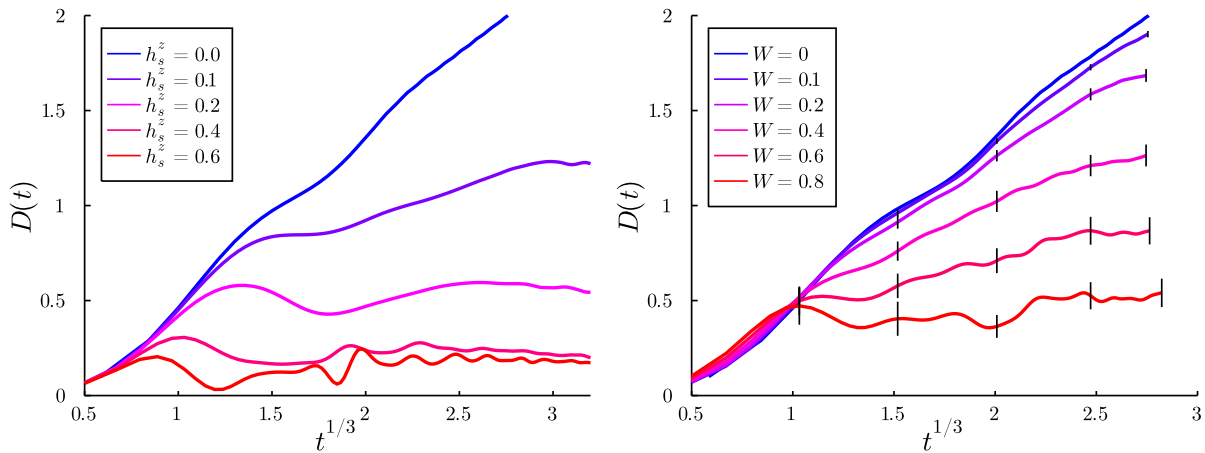

FIG. S3. **XXX Hamiltonian with symmetry breaking static perturbations** (left) $SU(2)$ symmetry is broken with staggered fields $H_1 = h_s^z \sum_j (-1)^j S_j^z$. (right) Translation symmetry is broken by adding randomness to the bond strengths - the perturbation is $H_1 = \sum_j \delta J_j S_j^z \cdot S_{j+1}^z$, where $\delta J_j$ is randomly drawn from $[-W, W]$. The plot shows the average $D(t)$ from 10 random realizations, with error bars indicating uncertainty in the average. All calculations were done with $\chi_{\max} = 1024$.

### D. Static perturbation evolutions

We also consider the isotropic Heisenberg spin chain $H_0$ with the addition of static perturbations $H_1$. Four choices of perturbations were considered, chosen to demonstrate the effects of perturbations that break or preserve $SU(2)$ symmetry and break or preserve translation symmetry. First, a staggered field was added to the XXX Hamiltonian to break $SU(2)$ symmetry. This resulted in a fairly quick crossover to diffusive behavior, as shown in Fig. S3. Second, we simulated the XXX hamiltonian with disordered bond couplings, preserving $SU(2)$ symmetry but breaking translation symmetry. Precisely, the XXX Hamiltonian was perturbed with $H_1 = \sum_j \delta J_j S_j^z \cdot S_{j+1}^z$, where $\delta J_j$ is randomly drawn from $[-W, W]$. Ten different disorder realizations were simulated — the average value of $D(t)$ over the different realizations is plotted in Fig S3. We see a slower crossover to diffusion. These calculations were done with both $\chi_{\max} = 512$ and $\chi_{\max} = 1024$, which agree on the shown time-frame.

Two perturbations that preserve the $SU(2)$ symmetry and (at least a subset of) the translation symmetry were also analyzed. The first of these was the addition of the next nearest neighbor couplings, $H_1 = J' \sum_j S_j \cdot S_{j+2}$. Finally, Heisenberg couplings were staggered by adding $H_1 = \delta J \sum_j (-1)^j S_j \cdot S_{j+1}$. As shown in the main text, the data from these perturbations is consistent with KPZ behavior, with any crossover to diffusion occurring at longer time scales than are accessible in the numerics. Fig S4 shows the evolutions for a variety of maximum bond dimensions $\chi_{\max}$, showing that $D(t)$ has converged for the shown times.

## II. BOLTZMANN EQUATION FOR NOISY PERTURBATIONS

We consider the unitary time evolution given by the Hamiltonian

$$H = H_0 + V(t), \tag{S5}$$

with the integrability-breaking perturbation $V(t) = \sum_n \eta_n(t) q_n^\sigma$, where $\eta_n$ is spatially correlated Gaussian noise with the correlation function $\langle \eta_n(t) \eta_{n'}(t') \rangle = \gamma \delta(t - t') f(n - n')$, and $\gamma$ is the strength of the noise. The function $f(n)$ is defined by a finite correlation length $\xi$ for the space correlation of the noise, e.g., $f(n) \sim e^{-|n|/\xi}$, but its precise functional form will not be important. The operator $q_n^\sigma$ is the density of one of the local conserved charges of the system. Since $H_0$ is integrable one can define the evolution of its local conserved charges $q_n^i$ via a Boltzmann equation, after averaging over the noise:

$$\partial_t \langle q_n^i(t) \rangle = I_i, \tag{S6}$$

where the scattering integral $I_i$ is a function of all conserved charges and clearly of the integrability breaking perturbation. The expectation values of the charges can be expressed in therms of the quasiparticle densities $\rho_s(\theta)$, where

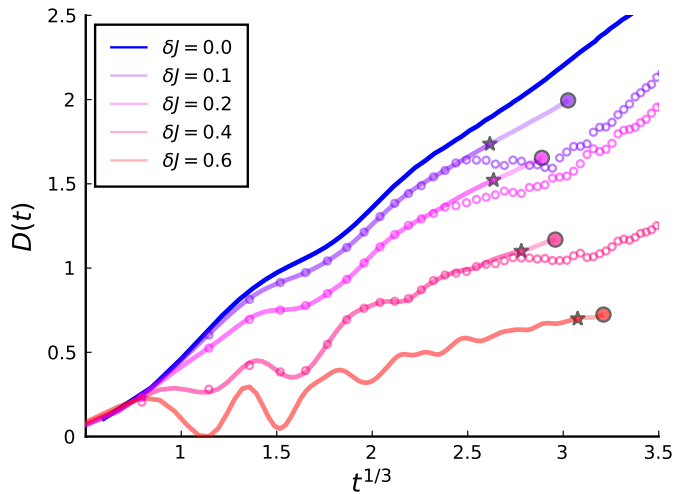

FIG. S4. **XXX Hamiltonian with staggered couplings** The curves show $D(t)$ computed with $\chi_{\max} = 2048$ (ending with the star), $\chi_{\max} = 1024$ (ending with the circle), and $\chi = 512$ (dotted lines). The data with the larger bond dimensions overlap completely, while the $\chi = 512$ data diverges for $t \gtrsim 15$.

the rapidity $\theta$ parameterizes their momenta $k_s(\theta)$, as

$$\langle q_i \rangle = \sum_s \int d\theta \; \rho_s(\theta) h_{i,s}(\theta), \tag{S7}$$

with single-particles eigenvalues $h_{i,s}(\theta)$. Here we have assumed that the background equilibrium state $|\rho\rangle$ is homogenous, and have dropped spatial derivatives. The Boltzmann equation can be recast into an equation for the evolution of the densities $\rho_s$. In the following we shall suppress the string index $s$ as we will work with diagonal scatterings among the same species. The equations are taken to hold separately for every $s$. Including only one particle-hole excitations (which is justified for smooth noise and small momentum exchange), where a quasiparticle with momentum $k(\theta)$ is scattered to a new momentum $k(\theta')$, the scattering integral can be written as [S10]

$$I_i = \int d\theta d\theta' \rho(\theta) \rho^h(\theta') \tilde{f}(\Delta k) \Delta q_i |\langle \rho | V | \rho, \theta \to \theta' \rangle|^2, \tag{S8}$$

where the distribution of holes is given by $\rho^h = k'/(2\pi) - \rho$, with $\rho^{\text{tot}} = k'/(2\pi)$ the total density of states. The charge difference $\Delta q_i$ is given by

$$\Delta q_i = [(1 + \partial(Fn))h_i](\theta') - [(1 + \partial(Fn))h_i](\theta), \tag{S9}$$

with the occupation numbers $n(\theta) = 2\pi\rho(\theta)/k(\theta)'$ and the shift function $F(\theta, \theta')$ defined in terms of an integral equation $F(\theta, \theta') = s(\theta, \theta') + \int d\alpha T(\theta, \alpha) F(\alpha, \theta')$, with $s(\theta, \theta')$ the scattering shift between two quasiparticles and $T(\theta, \theta')$ its derivative. The operator $(1 + \partial(Fn))$ acts as a linear operator in the space of functions where $1 \equiv \delta(\theta - \theta')$.

The function $\tilde{f}(\Delta k) \equiv \tilde{f}(k_s(\theta) - k_s(\theta'))$ is the Fourier transform of the noise variance $f(n)$ and it is defined up to shift of momenta of $2\pi$. We note an important subtlety: for the Heisenberg model, the Brillouin zone for $s$-strings is $(-\pi/s, \pi/s)$ whereas the noise has independent components in the microscopic Brillouin zone $(-\pi, \pi)$. Thus $\rho_s(k_s(\theta)) = \rho_s(k_s(\theta) + 2\pi/s)$—i.e., these are the same state—but two states $(s, \theta), (s, \theta')$ can be connected by noise components $\tilde{f}(k_s(\theta) - k_s(\theta') + 2\pi m/s)$ where $m = 0, \ldots, s - 1$ is an integer.

Taking this into account, the evolution of the densities is then given (reintroducing the index $s$) by

$$\partial_t \rho_s(\theta) = \int d\theta'' (1 + \partial(Fn))^T_{s;\theta,\theta''} \int d\theta' (\rho_s(\theta') \rho_s^h(\theta'') - \rho_s(\theta'') \rho_s^h(\theta'))$$
$$\times \left( \sum_m \tilde{f}(k_s(\theta) - k_s(\theta') + 2\pi m/s) \right) |\langle \rho | V | \rho, (s, \theta'') \to (s, \theta') \rangle|^2. \tag{S10}$$

Linearizing about an infinite temperature state where the the filling functions $n(\theta)$ are rapidity independent, we find

that they evolve as

$$
\partial_t n_s(\theta) = \frac{1}{2\pi} \int d\theta' k'_s(\theta') (n_s(\theta')(1 - n_s(\theta)) - n_s(\theta)(1 - n_s(\theta')))
$$
$$
\times \left( \sum_m \tilde{f}(k_s(\theta) - k_s(\theta') + 2\pi m/s) \right) |\langle \rho_s | V | \rho, (s,\theta) \to (s,\theta') \rangle|^2. \tag{S11}
$$

This equation has a simple, Fermi Golden Rule interpretation as a sum over in-scattering $\theta' \to \theta$ and out-scattering $\theta \to \theta'$ processes, weighted by the square of the matrix element of the perturbation and density of states factors. We can now approximate the matrix element by its low momentum limit [S13, S14]

$$
|\langle \rho | V | \rho, (s,\theta) \to (s,\theta') \rangle|^2 \simeq \lim_{\theta' \to \theta} |\langle \rho | V | \rho, (s,\theta) \to (s,\theta') \rangle|^2 = (h^{\mathrm{dr}}_{s,\sigma(\theta)})^2, \tag{S12}
$$

where the dressed single particle eigenvalue is defined as $h^{\mathrm{dr}} = (1 + Tn)^{-1}h$. This approximation is justified as follows: for small $s$, the function $\tilde{f}(\Delta k)$ is peaked around zero momentum transfer, relative to the size of the Brillouin zone; for large $s$, the momentum transfer is no longer small in this sense, but for quasiparticles with large string number $s$ all thermodynamic functions become flat in rapidity space.

We define the decay rate (inverse lifetime) of an $s$-string as the "out-scattering" contribution to the collision integral (S11), corresponding to processes out of the state $n_s(\theta)$:

$$
\Gamma_s \equiv -\frac{\partial_t n_s(\theta)}{n_s(\theta)} \bigg|_{\mathrm{out-scattering}} = (h^{\mathrm{dr}}_{\sigma,s}(\theta))^2 \frac{1}{2\pi} \int d\theta' k'_s(\theta')(1 - n_s) \sum_m \tilde{f}(k_s(\theta) - k_s(\theta') + 2\pi m/s). \tag{S13}
$$

Recall that the right-hand side is evaluated in the equilibrium state $|\rho\rangle$, which we take to be the infinite temperature state where $n_s \sim 1/s^2$ does not depend on rapidity. We are interested in the asymptotic behavior of the right hand side of the equation at large $s$, in order to extract the scaling in $s$ of the life-time of the quasiparticles. The noise spectrum in momentum space occupies a width $\sim 1/\xi$ where $\xi$ is the correlation length. Thus $\tilde{f}$ is approximately constant to this scale. If we keep $\xi$ fixed and take $s$ large, eventually the Brillouin zone of the $s$-string will be much smaller than $1/\xi$. In this regime one can replace $\tilde{f}$ by a constant, and truncate the sum over $m$ at the scale $1/\xi/(2\pi/s) \sim s/\xi$. This yields

$$
\Gamma_s \sim (h^{\mathrm{dr}}_{\sigma,s}(\theta))^2 \frac{s}{\xi} \int d\theta' k'_s(\theta')(1 - n_s(\theta')). \tag{S14}
$$

The rapidity integral scales as the size of the $s$-string Brillouin zone, $\sim 1/s$, so in the end we get

$$
\Gamma_s \sim (h^{\mathrm{dr}}_{\sigma,s})^2. \tag{S15}
$$

Note that we consider $\xi$ to be very large, so only low momentum exchanges are allowed, but finite, so $2\pi/s$ is can become much smaller than $1/\xi$ at large $s$. The opposite limit of $\xi \to \infty$ with $s$ large but finite leads to "momentum diffusion" [S10], and is quite different from our case where large strings can have their momentum (or rapidity) fully randomized after a single scattering event.

We now discuss the implications of Eq. (S15) for various choices of the charge $q$. First, for noise coupling to the spin density, $h^{\mathrm{dr}}_s \sim s^2$. Therefore large-$s$ strings are *short-lived* in this case, and as discussed in the main text we expect conventional diffusion. For noise coupling to the energy density, $h^{\mathrm{dr}}_s \sim 1/s$, so the lifetime of an $s$-string grows as $s^2$, as argued on more general grounds in the main text. This backs up our general argument in terms of Goldstone-mode scaling. Finally, the Boltzmann approach would also predict that if we instead choose the case of $q_n^\sigma = \vec{S}_n \cdot (\vec{S}_{n+1} \times \vec{S}_{n+2})$ (which is the next simplest conserved charge) we should find $(h^{\mathrm{dr}}_{\sigma,s}(\theta))^2 \sim s^{-6}$, suggesting a much smaller decay rate. The Boltzmann approach above would predict in this limit that quasiparticle decay should be subleading to screening, and KPZ scaling should be robust to noise. Numerical data for such a perturbation are, however, very similar to those for the energy density. Note however that our numerical simulations in this case were obtained for uncorrelated noise ($f(n) = \delta_{n,0}$), as correlated noise coupling to $q_n^\sigma = \vec{S}_n \cdot (\vec{S}_{n+1} \times \vec{S}_{n+2})$ would require dealing with MPOs of very large bond dimensions. We attribute such a discrepancy to higher particle-hole scattering processes (such as processes that are not diagonal in the $s$ index) that are not included in (S8), which become relevant if the noise is not smoothly varying in space. It would be important to clarify this issue in future works.

It would also be interesting to analyze the case of the weakly disordered random-bond XXX chain, which breaks translation-invariance, but preserves the $SU(2)$ symmetry. Formally, this corresponds to adding an energy-preserving delta function $\delta(\varepsilon_s(\theta') - \varepsilon_s(\theta))$ to the decay rate (S14). This restricts the accessible phase space to backscattering

processes, but apparently enhances the decay rate by a factor $1/|\varepsilon'|$ which suggests quickly decaying giant quasiparticles and ordinary diffusion. This analysis is consistent with our numerical results shown in Fig. S3. We leave a detailed analysis of disordered perturbations for future work, as they could reveal interesting connections with the localization properties of Goldstone modes [S15].

---



\end{document}